\documentclass[conference,letterpaper]{IEEEtran}
\IEEEoverridecommandlockouts
\usepackage{cite}
\usepackage{amsmath,amssymb,amsfonts}
\usepackage{graphicx}
\usepackage{textcomp}
\usepackage{xcolor}
\usepackage{url}
\def\BibTeX{{\rm B\kern-.05em{\sc i\kern-.025em b}\kern-.08em
    T\kern-.1667em\lower.7ex\hbox{E}\kern-.125emX}}
\usepackage{amsfonts, mathrsfs}
\usepackage{algorithmic}
\usepackage{array}
\usepackage{verbatim}
\usepackage{arydshln}
\usepackage{algorithm}
\usepackage{tabularray}
\usepackage{pifont}% http://ctan.org/pkg/pifont
\newcommand{\cmark}{\ding{51}}%
\newcommand{\xmark}{\ding{55}}%
\usepackage{listings}
\usepackage{amsthm}
\usepackage{booktabs,dcolumn}
\DeclareMathVersion{nxbold}
\SetSymbolFont{operators}{nxbold}{OT1}{cmr} {b}{n}
\SetSymbolFont{letters}  {nxbold}{OML}{cmm} {b}{it}
\SetSymbolFont{symbols}  {nxbold}{OMS}{cmsy}{b}{n}
\usepackage{multirow}
\usepackage{thmtools,thm-restate}
\usepackage{dcolumn}
\newcolumntype{d}[1]{D{.}{.}{#1}}

\usepackage{balance}
\usepackage{ragged2e}
\usepackage{makecell}

\DeclareMathOperator*{\argmin}{argmin}
\usepackage{tikz}
\usepackage{lipsum}
\usepackage{hyperref}

\begin{document}
\setlength{\columnsep}{0.24in}

\title{ResSR: A Computationally Efficient Residual Approach to Super-Resolving Multispectral Images\\
\thanks{This manuscript has been authored in part by UT-Battelle, LLC, under contract DE-AC05-00OR22725 with the US Department of Energy (DOE). The publisher acknowledges the US government license to provide public access under the DOE Public Access Plan (\href{https://energy.gov/doe-public-access-plan}{https://energy.gov/doe-public-access-plan}).}
}

\author{\IEEEauthorblockN{Haley Duba-Sullivan}
\IEEEauthorblockA{\textit{Department of Mathematics} \\
\textit{Purdue University}\\
West Lafayette, IN, USA \\
hduba@purdue.edu}
\and
\IEEEauthorblockN{Emma J. Reid}
\IEEEauthorblockA{\textit{National Security Sciences Directorate} \\
\textit{Oak Ridge National Laboratory} \\
Oak Ridge, TN, USA \\
reidej@ornl.gov}
\and
\IEEEauthorblockN{Sophie Voisin}
\IEEEauthorblockA{\textit{National Security Sciences Directorate} \\
\textit{Oak Ridge National Laboratory} \\
Oak Ridge, TN, USA \\
voisins@ornl.gov}
\and
\IEEEauthorblockN{Charles A. Bouman}
\IEEEauthorblockA{\textit{School of Electrical and Computer
Engineering} \\
\textit{Purdue University}\\
West Lafayette, IN, USA \\
bouman@purdue.edu}
\and
\IEEEauthorblockN{Gregery T. Buzzard}
\IEEEauthorblockA{\textit{Department of Mathematics} \\
\textit{Purdue University}\\
West Lafayette, IN, USA \\
buzzard@purdue.edu}
}

\maketitle

\begin{abstract}
Multispectral imaging (MSI) plays a critical role in material classification, environmental monitoring, and remote sensing. 
However, MSI sensors typically have wavelength-dependent resolution, which limits downstream analysis.
MSI super-resolution (MSI-SR) methods address this limitation by reconstructing all bands at a common high spatial resolution. 
Existing methods can achieve high reconstruction quality but often rely on spatially-coupled optimization or large learning-based models, leading to significant computational cost and limiting their use in large-scale or time-critical settings.

In this paper, we introduce ResSR, a computationally efficient, model-based MSI-SR method that achieves high-quality reconstruction without supervised training or spatially-coupled optimization.   
% Notably, ResSR decouples spectral and spatial processing into separate branches, which are then combined in a residual correction step.
Notably, ResSR decouples spectral and spatial processing into two sequential steps.
ResSR first computes a spectrally-informed high-resolution estimate of the MSI using singular value decomposition together with a spatially-decoupled approximate forward model. 
It then applies a residual correction step to restore low-frequency spatial consistency while preserving high-frequency detail recovered by the spectral reconstruction.
% The spectral branch uses singular value decomposition plus a spatially-decoupled approximate forward model to upsample the MSI, while the spatial branch uses bicubic upsampling.  The residual correction step combines these branches to recover accurate spectral and spatial MSI features.  
ResSR achieves comparable or improved reconstruction quality relative to existing MSI-SR methods while being 2$\times$ to 10$\times$ faster.
Code is available at \url{https://github.com/hdsullivan/ResSR}. 

\end{abstract}


\section{Introduction}

% Introduce the importance of high-spatial resolution multispectral imaging

% Introduce MSI-SR
\IEEEPARstart{M}{ultispectral} satellite sensors, such as MODIS, ASTER, VIIRS, Worldview-3, and Sentinel-2, generate multispectral images (MSI) containing dozens of bands, each acquired at a different optical wavelength.
These multispectral measurements enable analysis for applications where spectral signatures provide critical information beyond what is available in single-band imagery, such as material identification, environmental monitoring, and land-cover classification~\cite{bioucas2013hyperspectral}.
However, due to limitations of the optics and sensor hardware, MSI bands vary in spatial resolution \cite{kaufman1984atmospheric, lanaras_super-resolution_2017, lanaras_super-resolution_2018}.
This resolution mismatch results in inconsistent spatial detail across spectral bands, limiting accurate spectral analysis.
Consequently, MSI super-resolution (MSI-SR) algorithms are used to reconstruct MSIs at a common spatial resolution by super-resolving the lower-resolution bands.

Existing MSI-SR methods fall into two main categories. 
Deep learning-based
approaches~\cite{nguyen2025deep, sarmad2025difffusr, zheng_spectral_2022,
nguyen_sentinel-2_2021, salgueiro_romero_super-resolution_2020, lanaras_super-resolution_2018} learn a mapping from low- to high-resolution bands but require large supervised datasets and computationally expensive training. 
Model-based approaches use physical forward models paired with spatial regularization that couples neighboring pixels, often formulated using singular value decomposition (SVD)-based representations~\cite{lanaras_super-resolution_2017, ulfarsson_sentinel-2_2019, paris_hierarchical_2017}. 
Although effective, this model-based approach can lead to large, computationally expensive, spatially-coupled optimization problems. 
This motivates the development of more efficient MSI-SR methods that can exploit spectral correlations while reducing spatial coupling. 

\begin{table*}
    \centering
    \caption{Comparison of MSI-SR Methods}
    \vspace{-0.1cm}
    \label{tab:comp_methods}
    \begin{tabular}{|c|c|c|c|c|c|c|}
        \hline 
       \textbf{Method Name} & \multicolumn{1}{|p{1.8cm}|}{\centering \textbf{Handles 3+ resolutions}} &  \multicolumn{1}{|p{2.4cm}|}{\centering \textbf{Doesn't require\\supervised training}} & \multicolumn{1}{|p{2.8cm}|}{\centering \textbf{Doesn't require\\spatial regularization}} & \multicolumn{1}{|p{1.8cm}|}{\centering \textbf{Pixel-linear solution}} &\multicolumn{1}{|p{4cm}|}{\centering \textbf{Other comments}} \\
       \hline
        ResSR [ours] & \textcolor{green}{\cmark} & \textcolor{green}{\cmark} & \textcolor{green}{\cmark} & \textcolor{green}{\cmark} & Spectral regularization \\
        SupReME~\cite{lanaras_super-resolution_2017} &\textcolor{green}{\cmark} & \textcolor{green}{\cmark} & \textcolor{red}{\xmark} & \textcolor{red}{\xmark} &  Quadratic regularization \\
        LRTA~\cite{liu_hyperspectral_2021} & \textcolor{red}{\xmark} & \textcolor{green}{\cmark} & \textcolor{red}{\xmark} & \textcolor{red}{\xmark} & Spatial and spectral low-rank\\
        S2SHARP \cite{ulfarsson_sentinel-2_2019} & \textcolor{green}{\cmark}& \textcolor{green}{\cmark} & \textcolor{red}{\xmark} & \textcolor{red}{\xmark} & TV regularization\\
        SMUSH~\cite{paris_hierarchical_2017} & \textcolor{green}{\cmark} &\textcolor{green}{\cmark} & \textcolor{red}{\xmark} & \textcolor{red}{\xmark} & TV \& BM3D regularization\\
        DSen2~\cite{lanaras_super-resolution_2018} &\textcolor{green}{\cmark} & \textcolor{red}{\xmark} & \textcolor{green}{\cmark} & \textcolor{red}{\xmark} & Supervised CNN\\
       \hline
    \end{tabular}
    \vspace{-0.6cm}
\end{table*}

Table~\ref{tab:comp_methods} provides a high-level comparison of existing MSI-SR methods. 
For completeness, we also include an MSI-HSI fusion method, LRTA~\cite{liu_hyperspectral_2021}, and clarify its applicability to MSI-SR in Section~III.

% Introduce the novelties of ResSR 
In this paper, we propose ResSR, a computationally efficient, model-based MSI-SR method that achieves high-quality MSI reconstructions while eliminating spatially-coupled optimization. 
% Instead of solving a combined optimization, ResSR uses separate spectral and spatial branches computed independently and combined in a residual correction step.  
Instead of solving a combined optimization, ResSR decouples the spectral and spatial processing.  
First, ResSR applies a spectral reconstruction step using SVD and a spatially-decoupled approximate forward model.  
This step yields accurate high-frequency details, but can lead to low-frequency spectral shifts. 
To restore accurate low-frequency information, ResSR applies a residual correction step that corrects low-frequency distortion while preserving high-frequency structure recovered by the spectral reconstruction.
% The spatial branch uses bicubic upsampling, which maintains accurate low-frequency information but does not capture high-frequency details.  These two branches are combined using a residual correction process to recover accurate spectral and spatial MSI features.  

\begin{figure}
    \centering
    \includegraphics[width=0.46\textwidth]{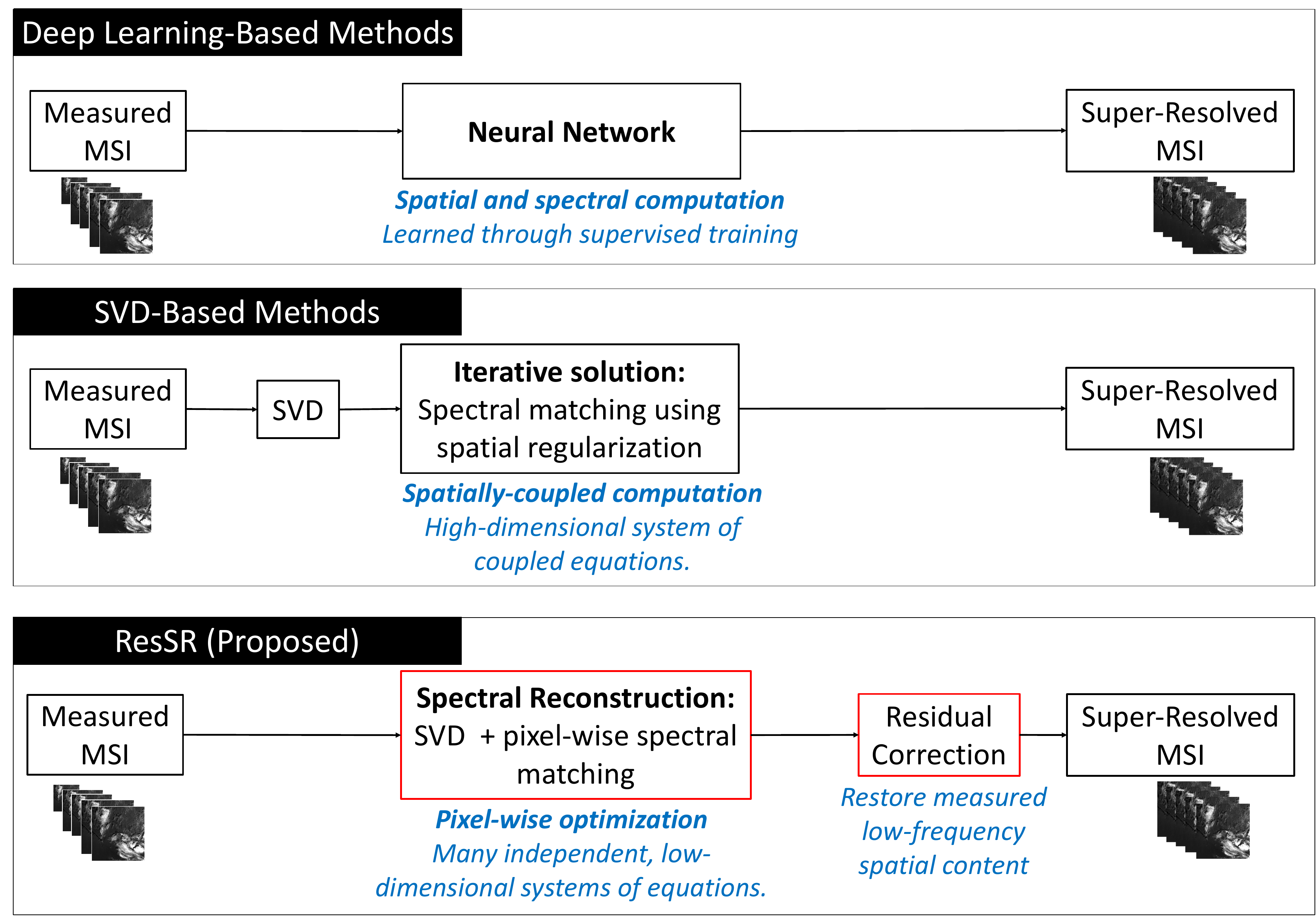}
    \vspace{-0.3cm}
    \caption{
    Comparison of the proposed ResSR pipeline with the standard pipeline for deep learning-based and iterative SVD model-based MSI-SR methods. ResSR decouples spatial and spectral processing, enabling a non-iterative, pixel-linear solution that significantly reduces computation while preserving accurate spatial detail.}
    \vspace{-0.5cm}
    \label{fig:comp_pipeline}
\end{figure}

Fig. \ref{fig:comp_pipeline} illustrates these decoupled steps in the ResSR pipeline in comparison to standard pipelines for deep learning-based and model-based MSI-SR methods.
The spectral step of ResSR is implemented as a collection of small, independent, per-pixel systems.
As a result, ResSR is ``pixel-linear,'' meaning that the computational complexity scales linearly with the number of pixels.

Our novel contributions include:
\begin{itemize}
        \item A computationally efficient and accurate MSI-SR method that decouples spatial and spectral processing, enabling a non-iterative, pixel-linear reconstruction without spatially-coupled regularization or supervised training,
        \item A lightweight residual correction strategy that restores measured low-frequency intensities while preserving high-spatial-frequency detail, compensating for the absence of explicit spatial regularization.
\end{itemize}
Experimental results on simulated and measured Sentinel-2 MSIs demonstrate that ResSR achieves reconstruction quality comparable to or exceeding existing MSI-SR methods while being 2$\times$ to 10$\times$ faster.

\section{ResSR Algorithm} \label{sec:ResSR}

\begin{figure*}[!t]
    \centering
    \vspace{0.05cm}
    \includegraphics[width=0.86 \textwidth]{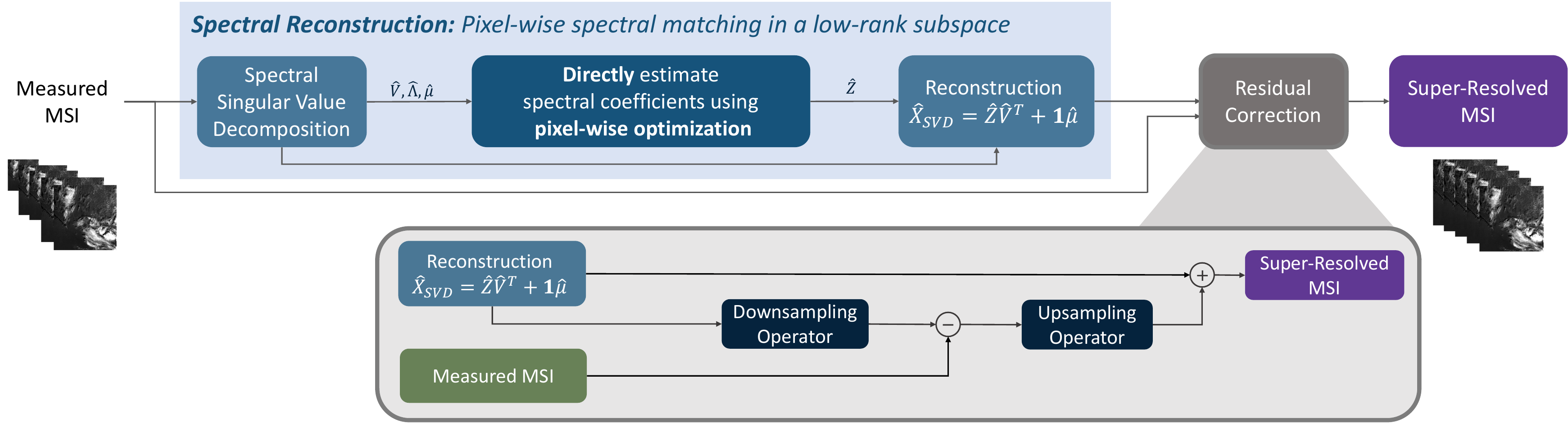}
    \vspace{-0.3cm}
    \caption{Overview of the ResSR pipeline.
    ResSR decouples spectral and spatial processing into two sequential steps.
    First, a spectral reconstruction step estimates a high-resolution MSI using SVD together with pixel-wise spectral matching in a low-rank subspace.
    Second, a residual correction step restores measured low-frequency spatial content by applying bicubic upsampling to the measurement residual.
    The final reconstruction preserves high-frequency detail from the spectral reconstruction while enforcing consistency with the measured data.
    % The spectral branch of ResSR uses singular value decomposition plus a spatially-decoupled approximate forward model to upsample the MSI, while the spatial branch uses bicubic upsampling. 
    % The residual correction process combines these two branches, extracting accurate high-frequency details from the spectral branch and low-frequency details from the spatial branch.  
    By decoupling the spatial and spectral processing, ResSR achieves computational efficiency throughout the entire method.
    }
\vspace{-0.6cm}
    \label{fig:ressr_pipeline}
\end{figure*}

Fig.~\ref{fig:ressr_pipeline} provides a visual depiction of the ResSR pipeline.  
ResSR consists of two sequential stages: (1) pixel-linear spectral matching in a low-rank subspace to form an initial high-resolution MSI estimate, and (2) residual correction that restores low-frequency consistency at each wavelength. The second stage refines the first-stage spectral estimate to yield the final super-resolved MSI.
% ResSR consists of two decoupled steps: (1) pixel-linear spectral matching in a low-rank subspace, and (2) residual correction using spatial upsampling.
% These branches are combined in the residual correction block to yield the super-resolved MSI.  

\subsection{Forward Model and Data Normalization}
\label{subsec:forward_model}

We denote the unknown high-resolution MSI that we wish to recover as $X\in \mathbb{R}^{N_p\times N_b}$, where $N_p$ is the number of pixels and $N_b$ is the number of bands.
More specifically, let
$$
X =[x_0, x_1, \dots, x_{N_b-1}] \ ,
$$
where each $x_i \in \mathbb{R}^{N_p}$ is a column vector representing the rasterized $i^{th}$ band of the super-resolved MSI.
In this reconstruction, each $x_i$ has the same resolution, which is measured in ground sampling distance (GSD) defined as the distance between pixel centers as measured on the ground.

The forward model that maps a reconstruction $X$ to sensor measurements $Y$ depends on the band, since different bands typically have different resolutions.
For band $i$, the sensor measurement is assumed to be
\begin{equation} 
\label{eq:forward_model} 
y_i = A_i x_i + \epsilon_i \ , 
\end{equation}
where $y_i \in \mathbb{R}^{\frac{N_p}{L_i^2}}$ is a column vector representing the rasterized $i^{\text{th}}$ band, $A_i$ is a linear operator that downsamples the MSI by a factor of $L_i$ over rows and columns, and $\epsilon_i \sim \mathcal{N}(0, \sigma^2 I)$ is independent additive white Gaussian noise. We assume that $A_i$ represents spatial block averaging over blocks of size $L_i \times L_i$ in the MSI.
Note that when $L_i = 1$, then $A_i = I$. 

{\bf Data normalization:} 
For numerical stability, each measured band is normalized prior to processing:
\begin{equation}
    y_{\text{norm}, i} = 
    \frac{y_i - p_2(y_i)}{p_{98}(y_i) - p_2(y_i)},
    \label{eq:normalization}
\end{equation}
where $p_2$ and $p_{98}$ denote the 2nd and 98th percentile intensities of
$y_i$.  For notational clarity, we use $y_i$ below to represent this normalized data, and we use this normalized data to compute a normalized estimate $\hat{x}_{\text{norm}, i}$ in
\eqref{eq:res_sr}. The final estimate in \eqref{eq:unnormalization} is obtained by inverting this normalization.

\subsection{Spectral Basis and Loss Function} \label{subsec: spectral basis}

Naively, the forward model in \eqref{eq:forward_model} suggests that our goal is to find $x_i$ that minimizes $\frac{1}{2\sigma^2}\|y_i - A_i x_i\|^2$.  
However, when $L_i>1$, the matrix $A_i$ is not invertible, so there is no unique solution for the corresponding $x_i$.
Moreover, treating each band separately does not exploit the fact that the MSI bands are correlated, and highly so for nearby frequency bands.

To exploit the correlation between bands, we assume that the MSI lies in a $K$-dimensional spectral subspace where $K<N_b$.
More specifically, we make an initial assumption (later subject to residual correction) that the MSI can be represented as
\begin{equation} 
\label{eq:subspace_rep}
    X_\text{SVD} = ZV^T + \mathbf{1} \mu,
\end{equation}
where $V$ is an $N_b \times K$ matrix of spectral basis vectors, $Z$ is an $N_p \times K$ image of basis coefficients, and $\mu \in \mathbb{R}^{1 \times N_b}$ is the spectral mean.
This type of subspace decomposition is known to represent MSIs accurately while reducing the dimensionality of the inverse problem~\cite{lanaras_super-resolution_2017, ulfarsson_sentinel-2_2019, paris_hierarchical_2017, liu_hyperspectral_2021}.

In addition to this spectral representation, we use a loss function for estimating $Z$ that includes pixel-wise spectral regularization and make an approximation to the forward model that eliminates the spatial coupling inherent in the downsampling operator $A_i$.

{\bf Spectral Basis:}
To estimate $\mu$ and $V$, we first perform a coarse interpolation of the low-resolution MSI bands so that all bands have the same resolution.
More specifically, we calculate
\begin{equation}
\label{eq:UpSampledMSI}
    \tilde{X} = [B_0 y_0, B_1 y_1, \dots, B_{N_b-1} y_{N_b-1}] \ ,
\end{equation}
where $B_i$ is bicubic interpolation by a factor of $L_i \times L_i$.

We then randomly subsample $N_s<<N_p$ pixels of the MSI to form a much smaller matrix $D\in \mathbb{R}^{N_s \times N_b}$ given by
\begin{equation}
    D = \text{subsample}_{N_s}\left( \tilde{X} \right)\ ,
\end{equation}
where $\text{subsample}_{N_s}(\cdot)$ randomly subsamples $N_s$ rows out of the original $N_p$ rows. 
From this, we estimate the row vector $\mu$ as the average of each column of $D$, given by
\begin{equation}
     \hat{\mu} = \frac{1}{N_s} \mathbf{1}^T D \ .
     \label{eq:mu_hat}
\end{equation}

We then estimate $V$ using the first $K$ right-singular vectors of the SVD of $D$ given by
\begin{equation}
    \hat{U}, \hat{\Lambda}, \hat{V} = \text{SVD}_{K} \left( D - \mathbf{1} \hat{\mu}\right) \ ,
    \label{eq:U_hat}
\end{equation}
where the columns of $\hat{V} \in \mathbb{R}^{N_b \times K }$ are the first $K$ orthonormal right-singular vectors and $\hat{\Lambda} \in \mathbb{R}^{K \times K}$ is the diagonal matrix of the corresponding singular values.
We use SVD instead of other techniques, such as non-negative matrix factorization, since our problem does not require non-negativity.

% Present data-fitting cost function
{\bf Loss Function:}
To estimate $Z$ using $\hat{\mu}$ and $\hat{V}$, we use a data-fitting term $f$ and a regularizing term $g$ (both defined below) to formulate a loss function
\begin{equation}  \label{eq:loss}
    \mbox{Loss}_\lambda (Z; \hat{V}, \hat{\Lambda}, \hat{\mu}) = f(Z; \hat{V}, \hat{\mu}) + \lambda g(Z; \hat{\Lambda}) \ ,
\end{equation}
where $\lambda$ is a user-selectable parameter that controls the regularization strength.

The data-fitting term, $f$, is a weighted sum of the negative log-likelihoods for the band-dependent forward models specified by~\eqref{eq:forward_model} and~\eqref{eq:subspace_rep} given by
\begin{align} 
\nonumber 
& f(Z; \hat{V}, \hat{\mu}) \\
\label{eq:data-fitting}
&= \frac{1}{2N_p\sigma^2} \sum_{i=0}^{N_b-1} \gamma_{L_i} L_i^2 \left \|y_i - A_i \left(Z \hat{V}^T+ \mathbf{1} \hat{\mu} \right)\mathcal{S}_i \right\|_2^2
\end{align}
where $\mathcal{S}_i \in \mathbb{R}^{N_b \times 1}$ selects the $i^{\text{th}}$ column of the matrix and $\gamma_{L_i}$ is a weighting vector whose elements are dependent on the resolution $L_i$. 

We specify the parameters $\gamma_{L_i}$ in terms of a single user-chosen parameter $\gamma_{HR} \in (0, 1)$ that describes the importance of the highest resolution bands on the reconstruction.
For any high-resolution band with $L_i=1$, we set $\gamma_{1} = \gamma_{HR}$.
For the remaining resolutions, we set 
\begin{equation}
\label{eq:gamma_explicit_form}
\gamma_{L_i} = \left( \frac{1 - \gamma_{HR}}{ \sum_{\ell \in {\cal L}} \frac{1}{\ell} }\right) \frac{1}{L_i} \ \text{ for }L_i\in {\cal L}\ ,
\end{equation}
where ${\cal L}$ is the set of lower resolutions in the MSI data.\footnote{For example, Sentinel-2 data includes bands at 10m, 20m, and 60m GSD, so ${\cal L}=\{2,6\}$.
}

By construction, the weights $\{\gamma_{L_i}\}$ sum to one across all resolutions, and the weights assigned to lower-resolution bands decrease inversely with their spatial resolution.

% Present regularization cost function
{\bf Regularization:}
The regularization term, $g$, in \eqref{eq:loss} is given by
\begin{equation}\label{eq:prior}
    g(Z; \hat{\Lambda} ) = \frac{1}{2N_pK} \|Z \hat{\Lambda}^{-1}\|_F^2.
\end{equation}
Intuitively, $g$ penalizes the spectral coefficients corresponding to less important right-singular vectors \cite{lang_simple_2020}; this penalty is achieved by using the truncated matrix of singular values $\hat{\Lambda}$ to inversely weight the spectral coefficients according to their energy in the SVD.
Note that the regularization in $g$ is pixel-wise since the Frobenius norm does not contain any spatial information. 

\subsection{Spatial Decoupling and Spectral Coefficients} 
% Present formal optimization problem
A direct approach using the loss function in \eqref{eq:loss} estimates $Z$ as
\begin{equation}
\label{eq:original_opt_problem} 
\hat{Z}_\text{coupled} = \argmin_{Z \in \mathbb{R}^{N_p \times K}} \mbox{Loss}_\lambda (Z; \hat{V}, \hat{\Lambda}, \hat{\mu} ) \ .
\end{equation}
However, taking the derivative of the loss with respect to $Z$ and setting it to 0 leads to a system of linear equations with spatial coupling introduced through multiplication by $A_i^T A_i$.
 
% Matching DC gain to estimate A^T A
To eliminate this spatial coupling, we note that $A_i^T A_i$ replaces each pixel in an $L_i \times L_i$ block with the average intensity of these pixels divided by $L_i^2$. For satellite multispectral imagery, which is typically smooth at the spatial scales corresponding to $L_i$, this operation is well approximated by multiplication by $L_i^{-2}$. 
More precisely, $\hat{Z} \hat{V}^T$ is a mean-subtracted estimate of the reconstructed MSI, which we assume varies smoothly at the spatial scales defined by the $L_i$.
Under this assumption, we approximate 
\begin{equation}
\label{eq:core_approximation}
A_i^T A_i \hat{Z} \hat{V}^T \approx \hat{Z} L_i^{-2} \hat{V}^T \ .
\end{equation}

Using this approximation, we eliminate the spatial-coupling term $A_i^T A_i$, in which case the first-order optimality condition for the approximate form of \eqref{eq:original_opt_problem} is 
\begin{multline}
\label{eq:approx_closed_form}
\hat{Z} \left(\sum_{i=0}^{N_b-1} \gamma_i \hat{V}_{i}^T \hat{V}_{i} + \frac{\lambda \sigma^2 }{K} \hat{\Lambda}^{-2} \right)\\ =  \sum_{i=0}^{N_b-1} \gamma_i L_i^2 A_i^T (y_i - \mathbf{1} \hat{\mu} \mathcal{S}_i) \hat{V}_i \ ,
\end{multline}
where $\hat{V}_{i} = \mathcal{S}_i^T \hat{V}$ is the $i$th row of $\hat{V}$ and we used the property $A_i \mathbf{1} = \mathbf{1}$ since $A_i$ is an averaging operator.

We solve \eqref{eq:approx_closed_form} to find the spectral coefficients $\hat{Z}$.  
Note that the $K \times K$ matrix to be inverted in \eqref{eq:approx_closed_form} is spatially-independent. Hence, finding $\hat{Z}$ reduces to solving $N_p$ small linear systems of size $K \times K$, where $K$ is small (2 in our experiments).

Although the resulting $\hat{Z}$ is an approximate solution to the optimization problem in~\eqref{eq:original_opt_problem}, it closely matches solutions obtained using iterative solvers such as ADMM.
Empirically, when applied to our simulated Sentinel-2 datasets (described in Section~III), the average NRMSE between reconstructions using the approximation and using ADMM range is 0.003 and differences are confined to high-frequency components.
In return, ResSR reduces compute time by 1000$\times$ relative to the ADMM solution.

% Give final estimated reconstruction
{\bf Spectral Reconstruction:}
From $\hat{Z}$, $\hat{V}$, and $\hat{\mu}$, we obtain the estimate of the super-resolved MSI from the spectral step of ResSR as
\begin{equation}
\label{eq:svd_sr}
    \hat{X}_\text{SVD} = \hat{Z} \hat{V}^T + \mathbf{1} \hat{\mu} \ .
\end{equation}

\subsection{Residual Correction} 
\label{subsec:residual_correction}
Since our spatially-decoupled reconstruction does not directly enforce spatial consistency, the spectral estimate in \eqref{eq:svd_sr} exhibits an inherent trade-off between spatial sharpness and intensity fidelity.
Large values of $\gamma_{HR}$ produce sharper detail but distort local intensity, while small values of $\gamma_{HR}$ preserve measured intensities at the expense of high-frequency structure.
To resolve this trade-off, we introduce a residual correction step that combines the measured low-frequency content with the high-frequency detail provided by the SVD-based estimate.

Recall that $A_i$ is downsampling by $L_i$ and that $B_i$ is bicubic upsampling by $L_i$. In practice, $B_iA_i$ acts as a low-pass filter and $I - B_iA_i$ extracts high-frequency content.
Using $I = B_iA_i + (I - B_iA_i)$ and $y_i \approx A_i \hat{x}_{\text{SVD},i}$, we may decompose
\begin{align}
\hat{x}_{\text{SVD}, i}
&= B_i A_i \hat{x}_{\text{SVD}, i} + (I - B_i A_i)\hat{x}_{\text{SVD}, i} \notag\\
&\approx B_i y_i + (I - B_i A_i)\hat{x}_{\text{SVD}, i}.
\end{align}
After rearranging the components on the right hand side, we obtain a residual form that we use to define the normalized estimate of band $i$: 
\begin{equation}\label{eq:res_sr}
    \hat{x}_{\text{norm}, i} = \hat{x}_{\text{SVD}, i} + B_i (y_i - A_i \hat{x}_{\text{SVD}, i}).  
\end{equation}
This residual correction restores the measured low-frequency intensity while retaining sharp high-frequency detail. 

As noted in Section~\ref{subsec:forward_model}, we invert the data normalization to obtain the final estimate of band $i$ as
\begin{equation}
    \hat{x}_{i}
    = (p_{98}(y_i) - p_2(y_i))\, \hat{x}_{\text{norm}, i} + p_2(y_i).
    \label{eq:unnormalization}
\end{equation}

We provide pseudocode for the entire ResSR algorithm in Algorithm~\ref{alg:res-sr}. 

\begin{algorithm}[ht] 
 \caption{ResSR}\label{alg:res-sr}
 \begin{algorithmic}
\renewcommand{\algorithmicrequire}{\textbf{Input:}}
 \REQUIRE $Y$, $\{A_i\}_{i=0}^{N_b-1}$, $\sigma$, $N_s$, $K$, $\gamma_{HR}$, $\lambda$ \\
  \vspace{2mm}
  \STATE \text{\textit{\# Normalize and subsample data}}
  \STATE Normalize $Y$ as in~\eqref{eq:normalization}
  \STATE $D \gets \text{subsample}_{N_s}\left( [B_1 y_1, B_2 y_2, \dots, B_{N_b} y_{N_b}] \right)$
  
  \vspace{2mm}
  \STATE \text{\textit{\# Compute SVD basis ($\hat{\mu}$, $\hat{V}$, and $\hat{\Lambda}$) }} 
  \STATE $\hat{\mu} \gets \frac{1}{N_s}  \mathbf{1}^T D$   
  \STATE $\hat{U}, \hat{\Lambda}, \hat{V} = \text{SVD}_{K} \left(D - \mathbf{1}{\hat{\mu}}\right)$ 
  
  \vspace{2mm}
  \STATE \text{\textit{\# Estimate basis coefficients ($\hat{Z}$)}} 
  \STATE $\hat{Z}$ from~\eqref{eq:approx_closed_form}
  
  \vspace{2mm}
  \STATE \text{\textit{\# Compute uncorrected super-resolved MSI ($\hat{X}_{\text{SVD}}$)}} 
  \STATE $\hat{X}_{\text{SVD}} \gets \hat{Z} \hat{V}^T + \mathbf{1} \hat{\mu}$
  
  \vspace{2mm}
  \STATE \text{\textit{\# Apply residual correction}} 
  \FOR {each band $i$}
   \STATE $\hat{x}_{\text{norm}, i} \gets \hat{x}_{\text{SVD}, i} + B_i (y_i - A_i \hat{x}_{\text{SVD}, i})$
  \ENDFOR

 \vspace{2mm}
 \STATE \text{\textit{\# Return super-resolved MSI ($\hat{X}$)}} 
 \STATE Unnormalize $\hat{x}$ as in~\eqref{eq:unnormalization}
 \RETURN $\{\hat{x}_i\}_{i=1}^{N_b}$
 \end{algorithmic}
 \end{algorithm}

\subsection{Computational Complexity}

An important practical contribution of ResSR is its computational efficiency. 
In SVD-based MSI-SR methods, the dominant cost is solving for the spectral representation coefficients. 
Typical model-based approaches lead to large, spatially-coupled systems with complexity up to $\mathcal{O}(K N_p^3)$ (or $\mathcal{O}(K N_p \log N_p)$ with Fourier acceleration).

In contrast, ResSR uses the diagonal approximation in
\eqref{eq:core_approximation} to convert this step into $N_p$ independent pixel-wise problems with only $K$ unknowns each. 
This yields a total complexity of $\mathcal{O}(N_p K^3)$, so ResSR is effectively linear in the number of pixels and embarrassingly parallel. Since $K$ is very small (e.g., $K=2$ in all our experiments), this leads to significant speedups while preserving accuracy.

\begin{table}
    \centering
    \caption{ResSR Parameters}
    \vspace{-0.2cm}
    \label{tab:res_sr_params}
    \begin{tabular}{c|c|c}
        
         Symbol & Description & Value Used \\
         \hline 
         $\sigma$ & Assumed noise in measured MSI & 0.02 \\ [1mm]
         $N_s$ & Number of subsampled pixels & $\sqrt{N_p}$\\[1mm]
         $K$ & Dimensionality of subspace & 2\\[1mm]
         $\gamma_{HR}$ & Impact of high-resolution bands on recon. & 0.99 \\[1mm]
         $\lambda$ & Spectral regularization weight & 0.5 \\
         \hline
    \end{tabular}
     \vspace{-.5cm}
\end{table}
 
\section{Experimental Results} \label{sec:exp_results}

\begin{table*}
\centering
\caption{NRMSE / SSIM for simulated datasets. We report the mean values over the $2\times$ and $6\times$ simulated Sentinel-2 datasets. \textcolor{blue}{\bf Best values are in blue} and second best values are in red. ResSR is competitive with DSen2 on the Sentinel-2 data used to train DSen2, and ResSR outperforms DSen2 on the out-of-distribution APEX data.}
\label{tab:nrmse_ssim_compact}
\renewcommand{\arraystretch}{1.25}
\scriptsize
\setlength{\tabcolsep}{3.5pt}

\begin{tabular}{c||c|c|c|c|c|c|c|c|c|c|c}
 & Method & Method Type & Runtime (s)
 & B1 & B5 & B6 & B7 & B8a & B9 & B11 & B12 \\
\hline
\multirow{4}{*}{\rotatebox[origin=c]{90}{APEX}}
 & ResSR & Model-based & \textbf{0.04}
 & \textcolor{red}{\textbf{0.179}} / \textcolor{red}{\textbf{0.917}}
 & \textcolor{red}{\textbf{0.058}} / \textcolor{red}{\textbf{0.987}}
 & \textcolor{blue}{\textbf{0.036}} / \textcolor{blue}{\textbf{0.992}}
 & \textcolor{blue}{\textbf{0.031}} / \textcolor{blue}{\textbf{0.994}}
 & \textcolor{blue}{\textbf{0.029}} / \textcolor{blue}{\textbf{0.995}}
 & \textcolor{blue}{\textbf{0.082}} / \textcolor{blue}{\textbf{0.942}}
 & \textcolor{red}{\textbf{0.077}} / \textcolor{red}{\textbf{0.952}}
 & \textcolor{red}{\textbf{0.096}} / \textcolor{red}{\textbf{0.967}} \\
 & SupReME & Model-based & 7.7
 & \textcolor{blue}{\textbf{0.098}} / \textcolor{blue}{\textbf{0.969}}
 & 0.067 / 0.982
 & 0.047 / 0.989
 & 0.034 / 0.991
 & 0.033 / 0.994
 & \textcolor{red}{\textbf{0.096}} / \textcolor{red}{\textbf{0.940}}
 & 0.119 / 0.928
 & 0.114 / 0.961 \\
 & LRTA & Model-based & 1.2
 & -- / --
 & \textcolor{blue}{\textbf{0.051}} / \textcolor{blue}{\textbf{0.989}}
 & 0.040 / 0.990
 & 0.034 / 0.993
 & 0.031 / 0.994
 & -- / --
 & 0.099 / 0.927
 & 0.134 / 0.943 \\
 & DSen2 & Neural Network & 1.3
 & 0.692 / 0.603
 & 0.095 / 0.968
 & \textcolor{red}{\textbf{0.037}} / \textcolor{red}{\textbf{0.991}}
 & \textcolor{red}{\textbf{0.032}} / \textcolor{red}{\textbf{0.993}}
 & \textcolor{red}{\textbf{0.030}} / \textcolor{red}{\textbf{0.995}}
 & 0.100 / 0.916
 & \textcolor{blue}{\textbf{0.043}} / \textcolor{blue}{\textbf{0.986}}
 & \textcolor{blue}{\textbf{0.060}} / \textcolor{blue}{\textbf{0.987}} \\
\hline

\multirow{4}{*}{\rotatebox[origin=c]{90}{Sentinel-2}}
 & ResSR & Model-based & \textbf{1.6}
 & \textcolor{red}{\textbf{0.045}} / \textcolor{red}{\textbf{0.896}}
 & \textcolor{red}{\textbf{0.013}} / \textcolor{red}{\textbf{0.987}}
 & \textcolor{red}{\textbf{0.014}} / \textcolor{red}{\textbf{0.986}}
 & \textcolor{red}{\textbf{0.015}} / \textcolor{red}{\textbf{0.985}}
 & \textcolor{red}{\textbf{0.015}} / \textcolor{red}{\textbf{0.984}}
 & \textcolor{blue}{\textbf{0.051}} / 0.843
 & \textcolor{red}{\textbf{0.020}} / \textcolor{red}{\textbf{0.974}}
 & \textcolor{red}{\textbf{0.020}} / \textcolor{red}{\textbf{0.976}} \\
 & SupReME & Model-based & 176.4
 & \textcolor{blue}{\textbf{0.043}} / \textcolor{blue}{\textbf{0.915}}
 & 0.015 / 0.986
 & 0.017 / 0.983
 & 0.017 / 0.984
 & 0.019 / 0.980
 & \textcolor{red}{\textbf{0.050}} / \textcolor{red}{\textbf{0.864}}
 & 0.027 / 0.973
 & 0.027 / 0.975 \\
 & LRTA & Model-based & 2.9
 & -- / --
 & \textcolor{blue}{\textbf{0.012}} / \textcolor{blue}{\textbf{0.988}}
 & 0.015 / 0.985
 & 0.016 / 0.984
 & 0.016 / 0.982
 & -- / --
 & 0.021 / 0.973
 & 0.022 / 0.975 \\
 & DSen2 & Neural Network & 4.2
 & 0.081 / 0.846
 & 0.014 / 0.985
 & \textcolor{blue}{\textbf{0.013}} / \textcolor{blue}{\textbf{0.989}}
 & \textcolor{blue}{\textbf{0.013}} / \textcolor{blue}{\textbf{0.989}}
 & \textcolor{blue}{\textbf{0.013}} / \textcolor{blue}{\textbf{0.988}}
 & 0.051 / \textcolor{blue}{\textbf{0.900}}
 & \textcolor{blue}{\textbf{0.012}} / \textcolor{blue}{\textbf{0.990}}
 & \textcolor{blue}{\textbf{0.015}} / \textcolor{blue}{\textbf{0.988}} \\

\hline
\end{tabular}
\vspace{-0.15cm}
\end{table*}

% Present comparison methods 
In this section, we  compare the performance of ResSR with several state-of-the-art MSI-SR methods.
Specifically, we compare to DSen2\footnote{DSen2 Code: github.com/lanha/DSen2}  \cite{lanaras_super-resolution_2018}, LRTA\footnote{Fixed Basis LRTA Code: my.ece.msstate.edu/faculty/fowler/software.html} \cite{liu_hyperspectral_2021}, and SupReME\footnote{SupReME Code: github.com/lanha/SupReME} \cite{lanaras_super-resolution_2017}. 
All methods use publicly available code from the authors with default parameters. 
Since we use bicubic interpolation in the residual correction step of ResSR, we also compare our results with bicubic interpolation. 
In the interest of space, we display bicubic interpolation results in only one experiment to establish its inability to capture high-frequency detail.

LRTA is excluded from our $6\times$ super-resolution experiments since it is an MSI-HSI fusion method rather than a pure MSI-SR approach.
MSI-HSI fusion combines a high-spatial-resolution, low-spectral-resolution MSI with a low-spatial-resolution, high-spectral-resolution hyperspectral image (HSI).
Although many fusion methods are closely related to MSI-SR~\cite{dong_model-guided_2021, simoes_convex_2015,
xu_iterative_2022, zhang_hyperspectral_2022, zhang_noise-resistant_2009,
li_fusing_2018, xu_hyperspectral_2023, dian2019hyperspectral,
zhang2018exploiting, dong2016hyperspectral}, they apply directly to MSI-SR only when the MSI contains exactly two spatial resolutions and the fusion model supports differing spectral ranges.
For completeness, we include LRTA~\cite{liu_hyperspectral_2021} in experiments when possible.

\begin{figure*}
    % \vspace{-2.8cm}
    \centering
    \includegraphics[width = 0.8\linewidth]{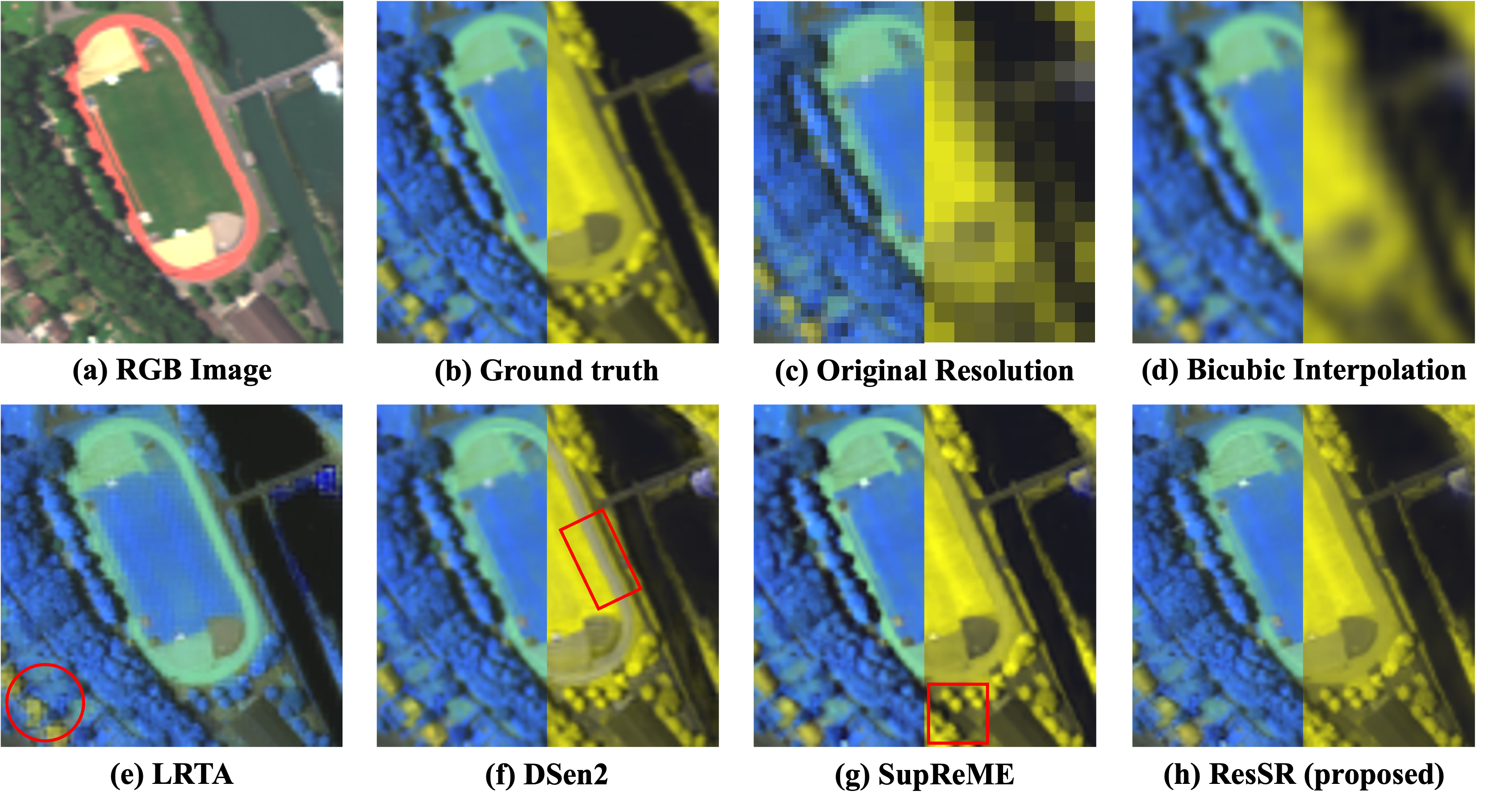}
    \vspace{-0.5cm}
    \caption{Comparison of $2\times$ and $6\times$ super-resolved bands to 2m GSD APEX ground truth and 4m / 12m GSD original resolutions, shown using a false-color composite. 
    In panels (d)-(h), the left half shows the $2\times$ super-resolution case (4m $\rightarrow$ 2m) using a false-color composite of bands B7, B11, B12, while the right half shows the $6\times$ super-resolution case (12m $\rightarrow$ 2m) using bands B1, B9, B9. 
    Since LRTA is not defined for multiple lower spatial resolutions, we exclude this method from our 6$\times$ super-resolution comparison.
    Bicubic interpolation produces blurred images, LRTA exhibits blocking artifacts, DSen2 introduces pixel-intensity distortions, and SupReME introduces high-frequency artifacts (shown in red).
    In contrast, ResSR produces sharp spatial detail with minimal artifacts, though it appears slightly sharper than the ground truth. Additionally, ResSR is roughly $2 \times$ faster than DSen2 and $100 \times$ faster than SupReME.}
    \vspace{-0.3cm}
\label{fig:apex_results}
\end{figure*}

\subsection{Data and Parameters}
In this section, we provide experimental details of the simulated and measured MSI data and algorithmic parameters used in experiments.
In all cases, downsampling was done using scikit-image's resize function with anti-aliasing enabled.
{\bf Data: } 
We use 3 simulated datasets and 1 measured dataset. 

The APEX simulated dataset is based on the APEX Open Science dataset acquired over Baden, Switzerland in June 2011~\cite{apex}.
The ground truth consists of 12 bands each at 2m GSD and can be downloaded from github.com/lanha/SupReME. 
We refer the reader to \cite{lanaras_super-resolution_2017} for more details on the ground truth simulation process.
We generated the MSI from the ground truth by downsampling the bands by a factor of 1, 2, and 6 to generate bands at 2m, 4m, and 12m GSD. 

The two simulated Sentinel-2 datasets are based on the measured Sentinel-2 dataset available from the Copernicus Open Access Hub service\footnote{Copernicus Open Access Hub: scihub.copernicus.eu}.  The MSIs in this dataset consist of 12 spectral bands with 10m, 20m, and 60m GSD \cite{noauthor_sentinel-2_nodate}. 
We disregarded Band 10 (B10) since it is primarily used for cirrus cloud detection and is very noisy \cite{lanaras_super-resolution_2017}.
We used the original data as ground truth and generated simulated MSI data by downsampling each band by a factor of 2 for the ``$2\times$~Sentinel-2'' dataset and a factor of 6 for the ``$6\times$~Sentinel-2'' dataset.
The $2\times$~Sentinel-2 dataset was used to quantitatively measure image reconstruction quality for B5, B6, B7, B8A, B11, and B12 and the $6\times$~Sentinel-2 dataset was used to quantitatively measure image quality for B1 and B9.

The simulated datasets have the advantage of allowing quantitative measures of reconstructed image quality since they include ground truth imagery.

Our measured dataset consists of 19 Sentinel-2 MSIs over various landscapes that we curated from the full Sentinel-2 data.
In this case, there is no ground truth.

{\bf Parameter Selection: }
Table~\ref{tab:res_sr_params} lists the parameters used in all experiments.
We chose $\sigma$ based on the estimated amount of noise present in measured Sentinel-2 data.
We chose $N_s$, $K$, and $\gamma_{HR}$ through tuning experiments on the APEX dataset.

\subsection{Results for Simulated Datasets}
Table~\ref{tab:nrmse_ssim_compact} reports the NRMSE, SSIM, and runtime for the simulated datasets, averaged over the $2\times$ and $6\times$ Sentinel-2 simulations.    
The best and second-best results for each band are shown in blue and red, respectively. 
Across the APEX data, ResSR achieves the best performance for bands B6–B9 and the second-best results for the remaining bands.  
This behavior is consistent with the strong spectral correlation between these bands and the 10m GSD measurements leveraged by ResSR. 
On simulated Sentinel-2 data, ResSR ranks first or second for all bands.
DSen2 performs best for Sentinel-2 B6–B9, which we attribute to the close similarity between our simulation pipeline and the training data used for DSen2.  
However, its performance degrades significantly on APEX bands B1 and B9, where the data distribution differs from its training set. 
DSen2 also underperforms on bands B11 and B12 for both datasets, likely because these SWIR bands lie far outside the 10m spectral range and are therefore more challenging to reconstruct \cite{lanaras_super-resolution_2017, lanaras_super-resolution_2018}.
In addition, DSen2 produces noticeably blurrier reconstructions in qualitative comparisons, likely due to reduced-resolution training.
LRTA attains the best score on band B5 but applies only to a single lower-resolution setting and incurs higher computational cost. SupReME achieves the best performance on APEX band B1 but is substantially slower, requiring on average $181\times$ the runtime of ResSR across image sizes.

Fig.~\ref{fig:apex_results} shows $2\times$ and $6\times$ reconstructions of APEX data using false-color composites (B7, B11, B12 for $2\times$; B1, B9, B9 for $6\times$).  
In each panel, the left half shows the $2\times$ super-resolution case (4m $\rightarrow$ 2m), while the right half shows the $6\times$ super-resolution case (12m $\rightarrow$ 2m). 
LRTA produces blocky artifacts at $2\times$ and is omitted from the $6\times$ comparison.  
DSen2 loses high-frequency spatial structure and introduces pixel-intensity distortions, while SupReME introduces high-frequency artifacts not present in the ground truth. 
In contrast, ResSR more closely matches ground-truth pixel intensities while preserving fine spatial detail.

\begin{figure}
    \centering
    % \vspace{-0.5cm}
    \includegraphics[width=0.83\linewidth]{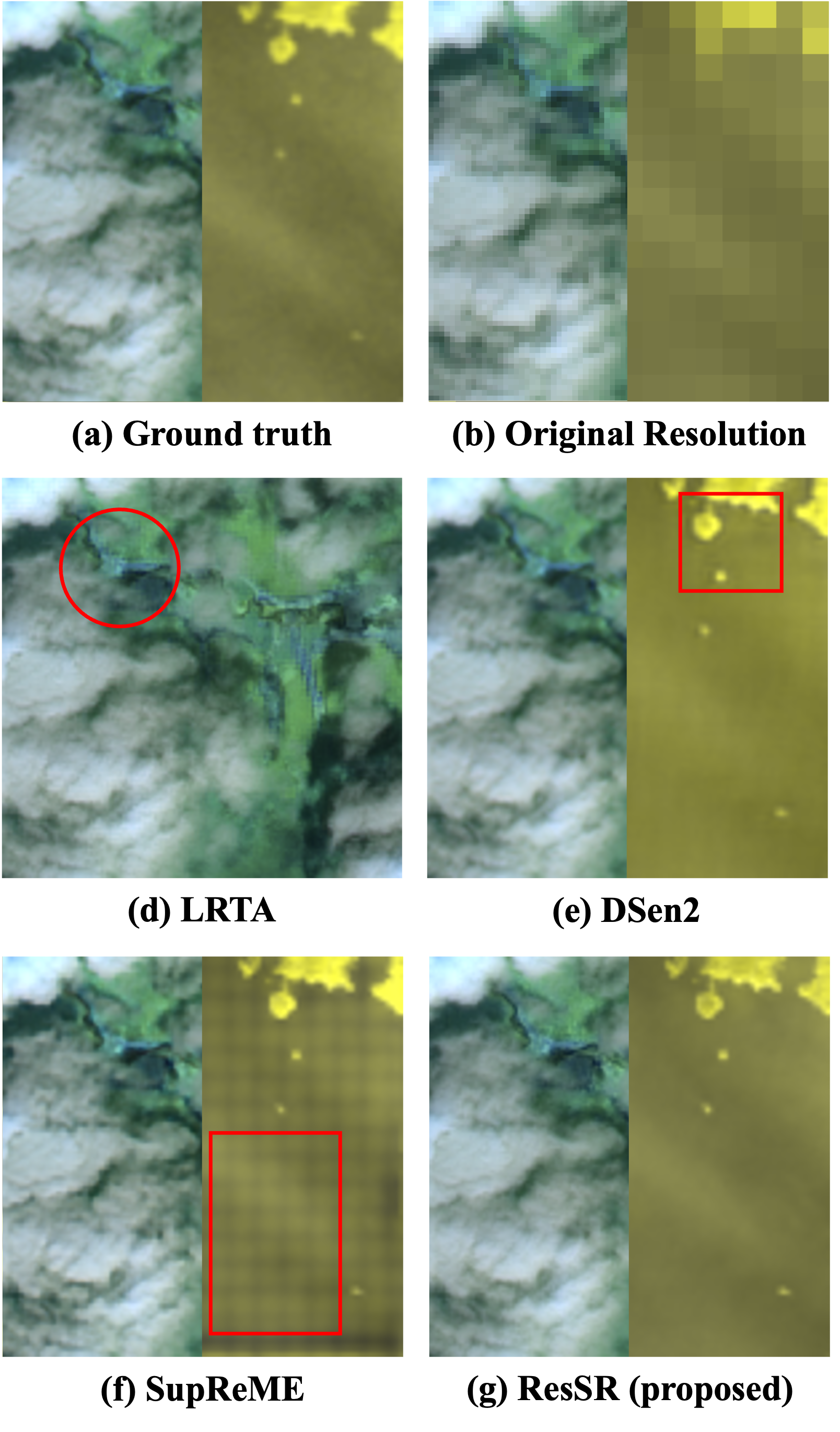}
    \vspace{-0.5cm}
    \caption{Comparison of $2\times$ and $6\times$ super-resolved bands to 20m GSD Sentinel-2 ground truth and 40m / 120m GSD original resolutions, shown using a false-color composite. 
    LRTA exhibits blocky artifacts, DSen2 loses high-frequency structure and introduces pixel-intensity distortions, and SupReME produces grid-like artifacts (shown in red). In contrast, ResSR matches the ground-truth pixel intensities and preserves fine spatial detail.}
    \label{fig:s2_sim_results}
    \vspace{-0.5cm}
\end{figure}

Fig.~\ref{fig:s2_sim_results} presents analogous results for simulated Sentinel-2 data.  
LRTA again introduces blocking artifacts at $2\times$ and is omitted from $6\times$.
DSen2 exhibits border and intensity distortions, and SupReME produces pronounced grid-like artifacts. ResSR consistently preserves high-frequency spatial structure, although mild over-sharpening is visible in some regions.
\subsection{Results for Measured Dataset} 
\label{subsec:measured_data_exp}

\begin{figure}
    % \vspace{-0.5cm}
    \centering
    \includegraphics[width=0.83\linewidth]{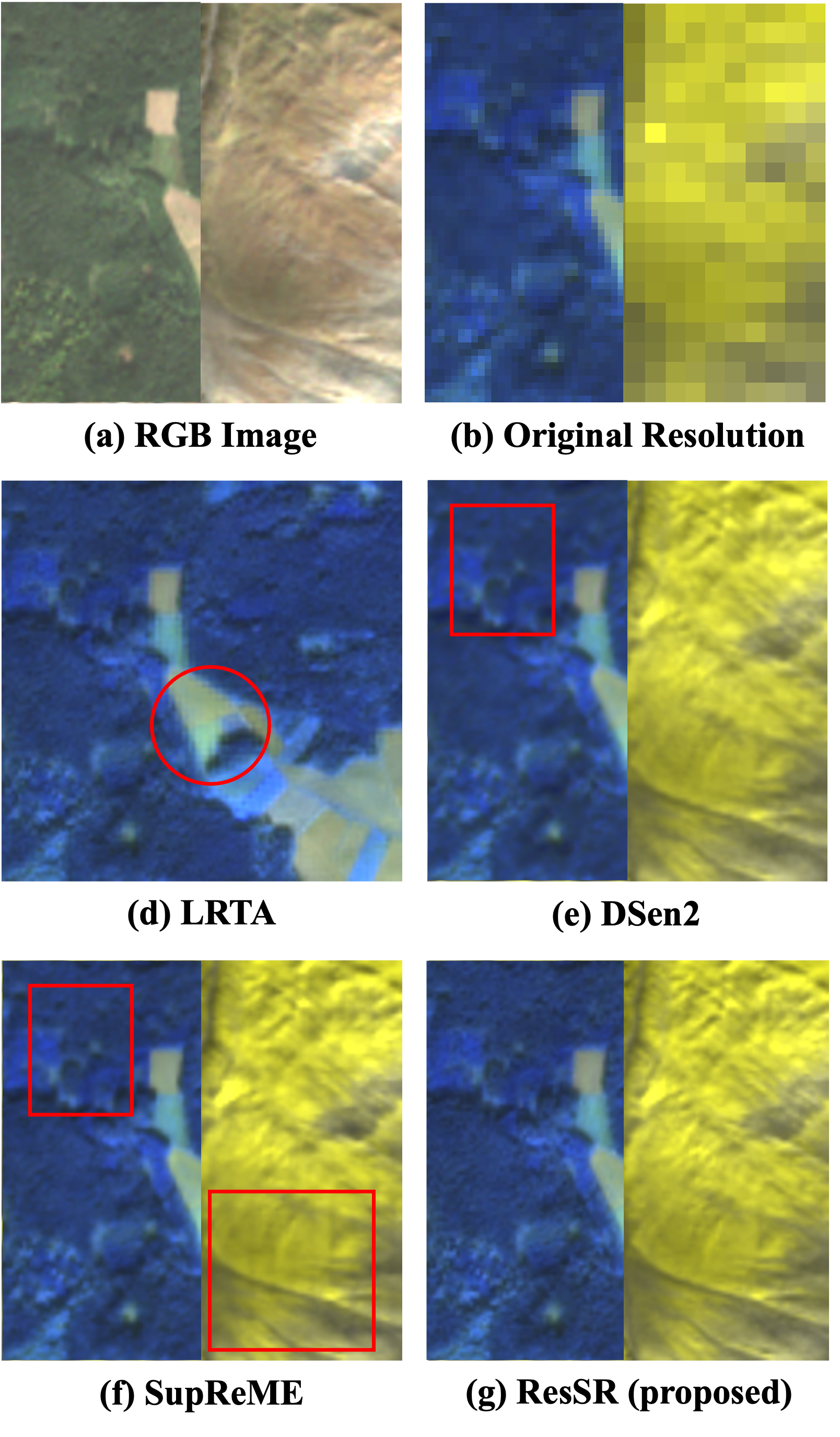}
    \vspace{-0.5cm}
    \caption{
    Comparison of $2\times$ and $6\times$ super-resolved bands to 20m / 60m GSD Sentinel-2 original resolutions, shown using a false-color composite. 
    LRTA introduces blocking artifacts, while DSen2 and SupReME lose high-frequency detail or introduce grid-like distortions (shown in red). In contrast, ResSR maintains accurate pixel intensities and restores spatial detail without visible artifacts.}
    \label{fig:s2_real_results}
    \vspace{-0.5cm}
\end{figure}

% ResSR creates sharper images with fewer artifacts/distortion in 60m real data
Fig.~\ref{fig:s2_real_results} shows $2\times$ and $6\times$ reconstructions of measured Sentinel-2 data using false-color composites (B7, B11, B12 for $2\times$; B1, B9, B9 for $6\times$).  
We also include an RGB image formed by the corresponding 10m GSD bands (B2, B3, and B4).
The LRTA reconstruction introduces blocking artifacts, especially noticeable within the red circle. 
DSen2 fails to recover fine spatial detail visible in the RGB reference image and introduces intensity distortions, especially noticeable in the forest region within the red rectangle. 
SupReME produces grid-like artifacts that are visually inconsistent with the reference imagery.
In contrast, ResSR preserves measured pixel intensities while restoring sharp high-spatial-frequency detail without introducing visible artifacts.

\subsection{Reconstruction Time} 
\label{sec:computation_speed}

\begin{table*}[!ht]
\centering
\caption{Reconstruction time for super-resolving all bands of a Sentinel-2 MSI with varying sizes. LRTA and DSen2 are run using a GPU with 16 GB of memory, while SupReME and ResSR are run using a CPU with 384 GB of memory. Note that LRTA does not super-resolve the two 60m bands. ``-'' indicates that the method ran out of memory. Results with shortest runtime are bolded.}\label{tab:time}

\newcolumntype{.}{D{.}{.}{-1}}
\makeatletter
\newcolumntype{B}[3]{>{\boldmath\DC@{#1}{#2}{#3}}c<{\DC@end}}
\newcolumntype{Z}[3]{>{\mathversion{nxbold}\DC@{#1}{#2}{#3}}c<{\DC@end}}
\makeatother
\begin{tabular}{c|c|c || .  .  .  .  .  . }
\hline
\multirow{2}{*}{Method} & \multicolumn{2}{c||}{\multirow{2}{*}{Method Type}}  & \multicolumn{6}{c}{Size of 10m GSD bands (in pixels)} \\

\cline{4-9} & \multicolumn{2}{c||}{} & \multicolumn{1}{c}{$180 \times 180$} & \multicolumn{1}{c}{$540 \times 540$} & \multicolumn{1}{c}{$1080 \times 1080$} & \multicolumn{1}{c}{$4380 \times 4380$} & \multicolumn{1}{c}{$7680 \times 7680$} &  \multicolumn{1}{c}{$10980 \times 10980$} \\
\hline
ResSR (ours) & Model-based & \multirow{2}{*}{\rotatebox[origin=c]{45}{CPU}} & \multicolumn{1}{B{.}{.}{-1}}{0.04\text{ sec}} & \multicolumn{1}{B{.}{.}{-1}}{0.4\text{ sec}} & \multicolumn{1}{B{.}{.}{-1}} {1.6\text{ sec}} & \multicolumn{1}{B{.}{.}{-1}}{23.8\text{ sec}} & \multicolumn{1}{B{.}{.}{-1}}{74.2\text{ sec}} & \multicolumn{1}{B{.}{.}{-1}}{155.8\text{ sec}}\\
SupReME & Model-based &  &  7.7 \text{ sec} & 43.9 \text{ sec} &176.4\text{ sec} & 4108.9\text{ sec} & 10472.2\text{ sec} & - \\
\hdashline
LRTA (20m only) & Model-based & \multirow{2}{*}{\rotatebox[origin=c]{45}{GPU}} &  1.2 \text{ sec} & 1.6 \text{ sec} & 2.9\text{ sec} & 28.4 \text{ sec}& - & - \\
DSen2 & Neural Network &  & 1.3 \text{ sec} & 2.0\text{ sec} & 4.2\text{ sec} & 44.7 \text{ sec} & 124.9 \text{ sec} & - \\
\hline
\end{tabular}
\vspace{-.5cm}
\end{table*}

% ResSR runs much faster!
Table~\ref{tab:time} shows the reconstruction time for super-resolving all 12 bands of a Sentinel-2 MSI with varying sizes. 
The shortest runtimes are in bold font.
DSen2 and LRTA are implemented in the Keras framework with TensorFlow as back-end, which automatically runs on a GPU, while SupReME and ResSR are implemented only for a CPU. 
For the reported comparison, we ran SupReME and ResSR on an Intel(R) Xeon(R) Gold 5122 CPU @ 3.60GHz, and we ran LRTA and DSen2 on an NVIDIA Quadro P5000 GPU. 
Note that the runtime of LRTA does not include super-resolution of the two 60m GSD bands. 

In all cases, ResSR has the shortest runtime.
For small MSIs of size $180\times180$, SupReME is 192.5$\times$ slower, LRTA is 30.0$\times$ slower, and DSen2 32.5$\times$ slower than ResSR. 
For larger MSIs of size $4380 \times 4380$, SupReME is 172.6$\times$ slower, LRTA 1.2$\times$ slower, and DSen2 1.9$\times$ slower than ResSR.
Note that the overhead associated with transferring data to the GPU likely contributes to the observed difference in runtime between small and larger MSIs for LRTA and DSen2.
In addition, SupReME, LRTA, and DSen2 all run out of memory for the largest MSI.
% and LRTA runs out of memory for the second largest MSI.
These results confirm that eliminating spatially-coupled optimization enables consistent speedups across image sizes.

\section{Conclusion}
\label{sec:conclusion}

In this paper, we introduced ResSR, a computationally efficient MSI-SR method that decouples spectral and spatial processing, generating high-quality reconstructions without spatially-coupled optimization or supervised training. 
ResSR first estimates a high-resolution MSI in a low-rank spectral subspace using SVD and a spatially-decoupled approximate forward model. It then applies a residual correction step based on bicubic upsampling of the measurement residual to restore measured low-frequency intensities while preserving the high-frequency detail recovered by the spectral reconstruction. This sequential design replaces spatially-coupled optimization with a lightweight, non-iterative algorithm, reducing the computational complexity from $\mathcal{O}(c N_p \log N_p)$ to $\mathcal{O}(c N_p)$, where $N_p$ is the number of pixels in the MSI and $c$ depends on the spectral subspace dimension.
ResSR naturally supports any number of distinct spatial resolutions, enabling straightforward application across a wide range of MSI sensors.

Experimental results on simulated and measured datasets demonstrate that ResSR consistently produces sharp reconstructions with minimal artifacts, achieving image quality comparable to or exceeding existing MSI-SR methods. 
Across all experiments, ResSR attained the lowest computational cost.
For $1080 \times 1080$ images, ResSR was $110\times$ faster than SupReME, $1.8\times$ faster than LRTA, and $2.6\times$ faster than DSen2.  ResSR was also able to reconstruct substantially larger images than any of the competing methods.  
These results indicate that ResSR enables high-fidelity MSI-SR at a computational cost low enough to support large-scale and time-critical applications, making efficient MSI-SR practical in settings where existing methods are prohibitive.

\bibliographystyle{IEEEtran}
\bibliography{main}
\end{document}